\documentclass[11pt,a4paper]{article}
\usepackage{jheppub-nosort,amsthm}
\usepackage{amsmath,amssymb,amsthm,amscd}
\usepackage{epsfig,comment}
\usepackage{graphicx,caption,subcaption}
\usepackage{yfonts,mathrsfs}
\usepackage{graphics,graphicx,epsfig}
\usepackage{amsfonts,psfrag,verbatim,color}
\newcommand{\be}{\begin{equation}}
\newcommand{\ee}{\end{equation}}
\newcommand{\beq}{\begin{equation}}
\newcommand{\eeq}{\end{equation}}
\newcommand{\ba}{\begin{eqnarray}}
\newcommand{\ea}{\end{eqnarray}}

\title{On AdS to dS transitions in higher-curvature gravity}
\author[a]{Xi\'an O. Camanho,}
\affiliation[a]{Max-Planck-Institut f\"ur Gravitationsphysik, Albert-Einstein-Institut, 14476 Golm, Germany}
\emailAdd{xian.camanho@aei.mpg.de}
\author[b]{Jos\'e D. Edelstein,}
\affiliation[b]{Department of Particle Physics and IGFAE, University of Santiago de Compostela\\ E-15782 Santiago de Compostela, Spain}
\emailAdd{jose.edelstein@usc.es}
\author[c]{Andr\'es Gomberoff,}
\affiliation[c]{Universidad Andres Bello, Departamento de Ciencias F\'\i sicas, Av. Rep\'ublica 252, Santiago, Chile}
\emailAdd{agomberoff@unab.cl}
\author[b]{J. An\'\i bal Sierra-Garc\'\i a}
\emailAdd{jesusanibal.sierra@usc.es}

\abstract{We study the possible existence of gravitational phase transitions from AdS to dS geometries in the context of higher-curvature gravities. We use Lanczos-Gauss-Bonnet (LGB) theory with a positive cosmological constant as a toy model. This theory has two maximally symmetric vacua with positive (dS) and negative (AdS) constant curvature. We show that a phase transition from the AdS vacuum to a dS black hole geometry takes place when the temperature reaches a critical value. The transition is produced by nucleation of bubbles of the new phase that expand afterwards. We claim that this phenomenon is not particular to the model under study, and shall also be part of generic gravitational theories with higher-curvature terms.}

\begin{document}
\maketitle

\section{Introduction}

The simultaneous existence of different AdS/dS vacua is quite a common feature in many gravitational theories. It is usually the result of coupling the metric to scalar fields \cite{scalar1,scalar2} or $p$-forms \cite{3form1,3form2,3form3,3form4,3form5}. This may give rise to a set of possible non-zero vacuum expectation values of the corresponding fields, contributing to the vacuum energy density or cosmological constant. Moreover, there are various mechanisms that may drive transitions between different vacua. These may be either a quantum tunneling process --an instanton \cite{instanton1,instanton2,instanton3}-- or a thermally activated transition \cite{thermal1,thermal2,thermal3,thermal4,thermal5}, possibly through thermalon mediation \cite{thermalon}. In most cases, these processes are directed from the high (dS) to the low (AdS) energy density vacua, minimizing energy as expected. There are instances, though, where the opposite is true, and the transition proceeds from an AdS to a dS geometry \cite{adsds,GuptSingh} (see, also, an earlier proposal in \cite{Odintsov1,Odintsov2}). In this article we would like to show a very simple scenario in which both vacua exist in the absence of any sort of matter, they are degenerate, and thermal effects may drive the system to undergo such a phase transition.

The existence of several vacua is in fact a well known feature of higher-curvature theories of gravity \cite{BD,Wheeler1,Wheeler2}. Among them, Lovelock theories play a prominent role since they yield second-order field equations (see \cite{bestiary,CESdS} for a recent review). Thereby, they provide a specially suitable and tractable playground that captures many important features that are rather generic in the context of higher-curvature gravities. Among them, the existence of new branches of black hole solutions, in correspondence with the set of vacua with different cosmological constants that pop out as soon as higher-curvature terms are brought into place.

We showed in \cite{CEGG1} that phase transitions among different vacua are possible and indeed generic \cite{CEGG2}. In analogy to the Hawking-Page process, above a critical temperature $T_c$ some of the Lovelock higher-curvature vacua shall decay by nucleating a bubble that hosts a black hole. The process is very similar to the thermalon mediated one discussed in \cite{thermalon}, in that bubbles are thermally nucleated. They separate space in two regions with different effective cosmological constants, in which black holes form in the interior. The thermalon in higher-curvature gravity differs from the one discussed in \cite{thermalon} in two important ways. On the one hand, the cosmological constant increases in the former at the end of the process. On the other, and most importantly, bubbles in higher-curvature theories are made of no matter, but of the gravitational field itself. After nucleation, dynamical instability of the bubble may trigger its expansion so that the interior phase will grow, ultimately filling the universe with the new cosmological constant. Sometimes, however, the instability may trigger its contraction, in which case the bubble would collapse. Naively, a naked singularity would form at the endpoint of the process. However, the contracting system becomes unstable well before that happens \cite{CEGG2}.

Even though the framework used in analyzing these transitions is completely general, previous works mostly focused on the case where both intervening branches where asymptotically AdS. This triggered the discussion about a suitable holographic interpretation of these gravitational phase transitions. Even though it is still unclear to us, we pointed out in \cite{CEGG1,CEGG2} that the whole process resembles that of quantum quenches in strongly coupled field theory (see, for instance, \cite{quenches}).

The present article is aimed at bridging one of the gaps in this line of research by considering the possibility of AdS to dS gravitational phase transitions. To that end we will consider the simplest non-trivial gravity possessing all the necessary ingredients, namely, the Lanczos-Gauss-Bonnet (LGB) theory with a positive {\it bare} cosmological constant.\footnote{We refer to the cosmological term in the action as {\it bare} to distinguish it from the effective cosmological constants giving the curvature of the different vacua; see (\ref{Lambdapm}).} The case involving a bare negative cosmological constant has been already analyzed in \cite{CEGG1}, where no AdS to dS transition was found: bubble configurations exist but they are all non-static; see \cite{CEGG2}. In the canonical ensemble, whether or not the transition takes place can be decided by evaluating and comparing the Euclidean action among the various smooth classical configurations satisfying the appropriate boundary conditions. This same on-shell Euclidean action can be used to compute the probability of bubble formation at a given temperature. We discuss the dependence of the critical temperature --below which the AdS vacuum is the thermodynamically preferred configuration-- on the LGB coupling constant, $\lambda$. Even though for small temperatures the AdS vacuum seems stable, it is actually metastable. There is always a probability for a bubble to be nucleated.

In Lorentzian signature, the dynamics of the bubble can be analyzed. Once produced, it will expand and the dS vacuum will take over changing the asymptotics of the spacetime in finite proper time. This entails a quite interesting scenario in which thermal AdS can decay into a black hole dS geometry. At least two novel features occur, in comparison to the AdS to AdS phase transitions studied earlier in \cite{CEGG1,CEGG2}. The thermalon does not exist above a threshold temperature $T_\star$ that is a function of $\lambda$. In particular, this implies that there is a critical value, $\lambda_\star$, above which the thermalon free energy is always positive. Moreover, when the bubble expands, it ends up reaching the cosmological horizon in finite proper time, which prevents the possibility of choosing reflecting boundary conditions as in the case of AdS (even though the starting geometry {\it is} AdS).

\section{Thermalon in quadratic (LGB) gravity with $\Lambda > 0$}

The LGB theory of gravity is given by the following action
\begin{equation}
\mathcal{I} = \frac{1}{16\pi G} \int d^{d}x\, \sqrt{-g}\,\left( R - \frac{(d-1)(d-2)}{L^2} + \frac{\lambda L^2}{(d-3)(d-4)}\mathcal{R}^2 \right)-\mathcal{I}_\partial ~,
\label{action}
\end{equation}
where $\mathcal{R}^2$ amounts to the combination, $\mathcal{R}^2 = R^{2} - 4 R_{\mu\nu} R^{\mu\nu} + R_{\mu\nu\lambda\rho} R^{\mu\nu\lambda\rho}$, which guarantees that the equations of motion are second order in $d\geq 5$ spacetime dimensions. Notice that the bare cosmological constant is positive, $\Lambda = (d-1)(d-2)/2 L^2 > 0$, and we have a single dimensionless coupling, $\lambda$, while dimension-full parameters are the Newton constant and the characteristic length, $L$. We take $\lambda > 0$ which is the natural sign inherited, for instance, from string theory embeddings of (\ref{action}). The additional boundary term, $\mathcal{I}_\partial$, is necessary in order to have a well defined variational principle (see \cite{CEGG1} for further details). It plays a fundamental role in the phase transitions discussed in this letter, as we will explain below. 

\begin{figure}[ht]
\begin{center}
\includegraphics[width=0.49\textwidth]{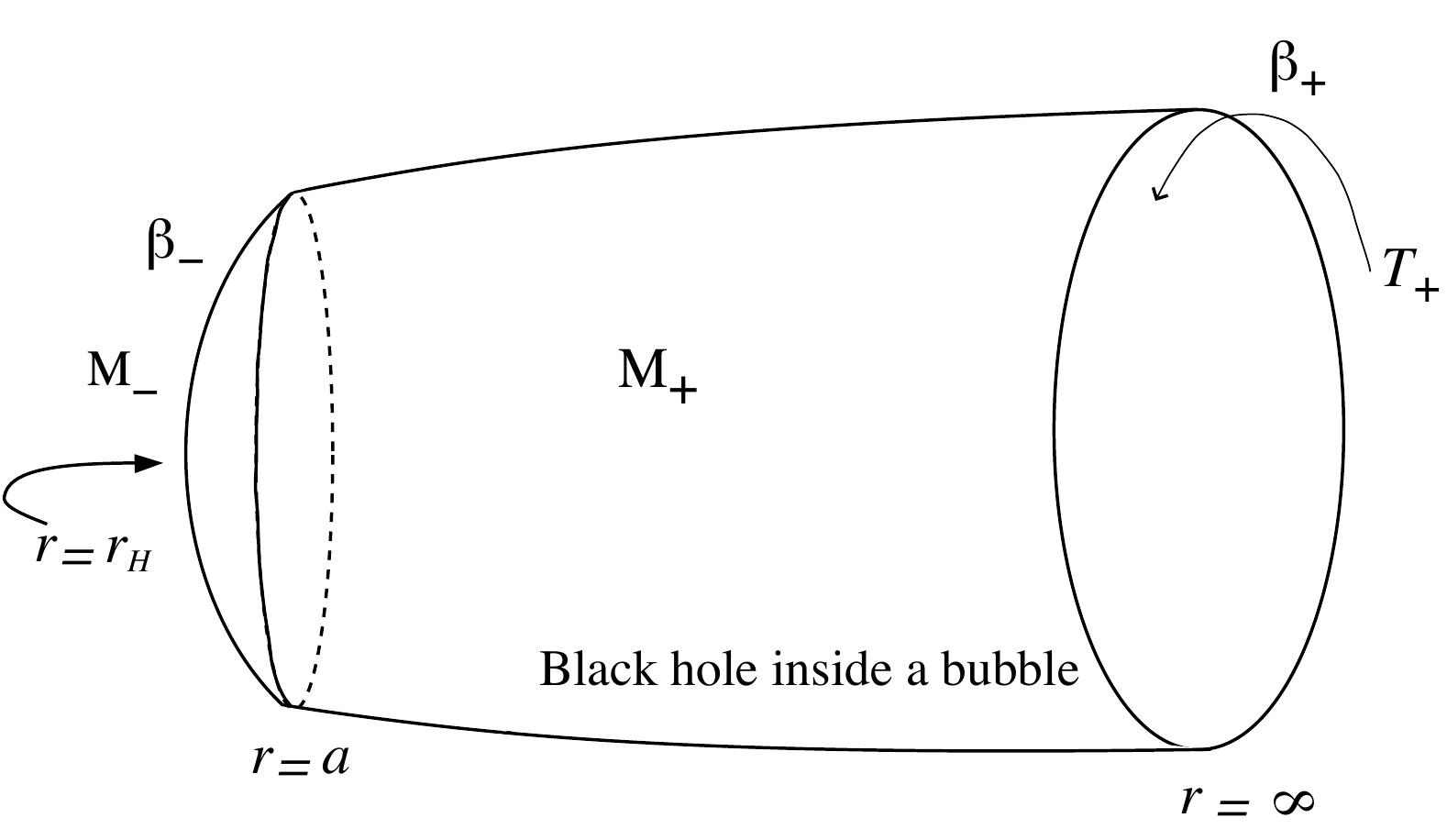}
\end{center}
\caption{The thermalon: Euclidean section for the bubble hosting a (negative branch; $f_-$ in (\ref{GBbranches})) black hole in its interior with (positive branch; $f_+$ in (\ref{GBbranches})) AdS asymptotics.}
\label{thermalonfig}
\end{figure}

Black hole solutions for this theory can be readily obtained by means of a simple spherically symmetric ansatz of the form
\begin{equation}
ds^2 = -f(r)\,dt^2 + \frac{dr^2}{f(r)} + r^2 d\Omega^2_{d-2} ~,
\label{BHansatz}
\end{equation}
where $d\Omega^2_{d-2}$ is the metric of a unit round sphere. The equations of motion can be readily solved and two branches of solutions emerge \cite{BD,Wheeler1,Wheeler2}:
\begin{equation}
f_{\pm}(r) = 1 + \frac{r^2}{2\lambda L^{2}} \left( 1\pm \sqrt{1 + 4\lambda \left( 1 + \frac{{\rm M}_\pm}{r^{d-1}}\right)} \right) ~,
\label{GBbranches}
\end{equation}
where M$_\pm$ is the (properly normalized) mass parameter of the spacetime \cite{BD}. The dS case under consideration has been analyzed in detail in \cite{CaiGuo}. Notice that the only branch admitting a black hole solution ({\it i.e.}, a smooth event horizon) is that with the minus sign. Each of the two branches in (\ref{GBbranches}) is associated with a different value of the effective cosmological constant,
\begin{equation}
\Lambda_{\pm} = - \frac{1\pm \sqrt{1 + 4\lambda}}{2\lambda L^{2}} ~.
\label{Lambdapm}
\end{equation}
It is immediate to see that $\Lambda_-$ is positive, while the negative $\Lambda_+$ is afflicted by the Boulware-Deser (BD) instability \cite{BD}. We shall consider in what follows Euclidean static bubble solutions ({\it i.e.}, thermalons) with either branches in the outside/inside. The only possibility that entails a regular Euclidean section --to be considered as a saddle point of the path integral-- is that hosting a negative branch solution in the interior of the bubble, since only that branch displays event horizons cloaking singularities (see Figure \ref{thermalonfig}).
This means that the asymptotics fixing the boundary conditions in the path integral is that of the positive branch; {\it i.e.}, we deal with asymptotically AdS geometries.  Notice that the reversed configuration is singular. Therefore, a gravitational phase transition mediated by thermalon nucleation from dS to AdS is forbidden in LGB theory with $\Lambda > 0$.

In addition to being required for the variational principle to be well defined, boundary terms play a fundamental role in the description of \textit{weak} or distributional solutions to the equations of motion. When considering a co-dimension one surface, $\Sigma$, splitting the spacetime manifold in two, it is convenient to rewrite the action also as two bulk and one surface contributions, according to the spacetime splitting. We may then write (see \cite{CEGG2} for details), 
\begin{equation}
\mathcal{I} = \mathcal{I}^- + \mathcal{I}_\Sigma +\mathcal{I}^+ - \mathcal{I}_\partial ~,
\label{fullaction}
\end{equation}
where the bulk contributions, $\mathcal{I}^{\pm}$, have the same form as (\ref{action}), whereas the bubble term can be in turn split into two terms, each one corresponding to either side. Both are formally the same as the original boundary term, $\mathcal{I}_\partial$, except for the fact that they are evaluated at the position of the bubble,
\begin{equation}
\mathcal{I}_\Sigma = \mathcal{I}_\partial^+ - \mathcal{I}_\partial^- ~.
\label{Ibubble}
\end{equation}
This action is well adapted to the problem we are interested in as now the variation of the bulk terms still results in the usual Lovelock field equations, whereas the variation of the surface terms yields the junction conditions along $\Sigma$. The latter amount to continuity conditions on the canonical momenta of the theory, 
\begin{equation}
\Pi^+_{ij} = \Pi^-_{ij} ~.
\label{equalsPi}
\end{equation}
We are interested in non-singular static configurations as the one displayed in Figure \ref{thermalonfig}. Being static, the Euclidean sections have the form
\begin{equation}
ds^2 = f_{\pm}(r_\pm) dt_\pm^2 + \frac{dr_\pm^2}{f_{\pm}(r_\pm)} + r_\pm^2 d\Omega_{d-2}^2 ~,
\label{Eucsection}
\end{equation}
where the $\pm$ signs correspond to the outer/inner regions. Projecting from either side of the bubble onto the timelike hypersurface $\Sigma$, which we define parametrically as $t_\pm=T_\pm(\tau)$, $r_+=r_-=a(\tau)$, we get
\begin{equation}
ds_\Sigma^2 = d\tau^2 + a^2(\tau)\,d\Omega_{d-2}^2 ~,
\label{inducedmetric}
\end{equation}
where the radial coordinate has to be continuous across the junction. For the orientation we are interested in --gluing the interior dS black hole with the exterior AdS asymptotics--, we can just take a single-valued radial coordinate $r = r_\pm$. Furthermore, we have chosen
\begin{equation}
f_\pm\dot{T}_\pm^2 + \frac{\dot{a}^2}{f_\pm} = 1 ~,
\label{fpm}
\end{equation}
ensuring that the physical length of the time circle is the same as seen from both sides. In particular, in the case of a static bubble, $a=a_\star$, it relates the periodicities, $\beta_\pm$, between the inner and outer Euclidean time coordinates, $\sqrt{f_+(a_\star)}\,\beta_+ = \sqrt{f_-(a_\star)}\,\beta_-$. This condition ensures the continuity of the metric in the sense that there exists a change of coordinates relating the inner/outer metrics at the bubble location.

The junction conditions (\ref{equalsPi}) have just diagonal components related by a conservation equation that constrains them in such a way that only the $\tau\tau$ component matters,
\begin{equation}
\Pi_{\tau\tau}^+(a,\dot{a}) = \Pi_{\tau\tau}^-(a,\dot{a}) ~.
\label{Pitautau}
\end{equation}
The remaining components correspond to the $\tau$ derivative of the former \cite{Davis2003,wormholes},
\begin{equation}
\frac{d}{d\tau}\left(a^{d-2}\, \Pi^\pm_{\tau\tau}(a,\dot{a})\right) = (d-2)\, a^2 \dot{a}\, \Pi^\pm_{ii}(a,\dot{a}) ~.
\label{Bianchi}
\end{equation}
The momenta $\Pi_{\tau\tau}^{\pm}(a,\dot{a})$ are computed in terms of the extrinsic and intrinsic curvatures of the bubble (see \cite{CEGG2} for a detailed discussion), yielding, in Lorentzian signature,
\begin{equation}
\Pi_{\tau\tau}^{\pm}(a,\dot{a}) = \frac{\sqrt{\dot{a}^2+f_\pm(a)}}{a} \int_0^1\! d\xi \left[ 1 + 2\lambda \frac{L^2}{a^2} \left( \dot{a}^2 + 1 - \xi^2 (f_\pm(a) + \dot{a}^2) \right) \right] ~.
\end{equation}
When these are plugged into (\ref{Pitautau}), the junction condition can be easily seen to admit a more intuitive expression in terms of a potential, $V_{\rm th}(a)$,
\begin{equation}
\frac12\dot{a}^2 + V_{\rm th}(a) = 0 ~,
\label{dotapot}
\end{equation}
from which the transverse components of the junction conditions --see (\ref{Bianchi})-- just amount to $\ddot{a}=-V_{\rm th}'(a)$. We can work out explicitly the form of $V_{\rm th}(a)$,
\begin{equation}
V_{\rm th}(a) = \frac{1+4\lambda}{24\lambda} \left[ a^{d-1} \frac{f_-(a) - f_+(a)}{{\rm M}_+-{\rm M}_-} + \frac{4\lambda}{1+4\lambda} \frac{{\rm M}_- f_-(a) - {\rm M}_+ f_+(a)}{{\rm M}_+-{\rm M}_-} + \frac{8 (a^2 + 2\lambda)}{1+4\lambda} \right] ~.
\label{V2}
\end{equation}
The bubble dynamics is governed by the Minkowskian version of the junction conditions. The thermalon is nothing but its Euclidean static solution, $a(\tau) = a_\star$, with $V_{\rm th}(a_\star) = V_{\rm th}'(a_\star) = 0$. It should be noticed that even though the thermalon possesses five parameters (M$_\pm$, $\beta_\pm$ and $a_\star$), there are four equations relating them: two coming from the junction conditions, one from the matching of the thermal circles at the bubble location and one from the usual Hawking condition to avoid deficit angles at the black hole event horizon, fixing $\beta_-$. All in all, the configuration has a single free parameter that, in the canonical ensemble, is taken to be the inverse temperature measured by an asymptotic observer, $\beta_+$. All physical quantities can be written in terms of it. The general expressions are not particularly enlightening.

\section{Gravitational AdS to dS phase transition}

The usual thermodynamic picture of the black holes dynamics can be readily described in a semiclassical approach to the quantization of gravity. We can formally define the canonical ensemble at temperature $1/\beta_+$ as given by the path integral over all asymptotically AdS metrics  identified in Euclidean time with period $\beta_+$,
\begin{equation}
\mathcal{Z} = \int\! \mathcal{D}g\ {e^{-\hat{\mathcal{I}}[g]}}\approx e^{-\hat{\mathcal{I}}[g_{cl}]} ~, \qquad \hat{\mathcal{I}}=-i\mathcal{I} ~.
\label{path}
\end{equation}
The dominant contributions come from the saddle points, \textit{i.e.} classical solutions of the field equations, $g_{cl}$. The simplest saddles correspond to static manifolds that can be trivially rotated back and forth between their Lorentzian and Euclidean sections. 

As explained previously, the action in this case has several bulk and surface contributions that have to be all taken into account. Despite their complicated form, it turns out that, once the junction conditions are imposed, the Euclidean action has the expected form of a free energy (multiplied by the inverse temperature)
\begin{equation}
\hat{\mathcal{I}} = \beta_+ F = \beta_+ M_+ - S ~,
\end{equation}
generalizing the analogous result that originated the thermodynamic approach to black holes. The inverse temperature $\beta_+$ corresponds to the periodicity in the outer time coordinate whereas the entropy is unchanged by the presence of the bubble \cite{CEGG2}. The relation between the various quantities is such that the first law of thermodynamics is also verified.

\begin{figure}[ht]
\begin{center}
\includegraphics[width=0.58\textwidth]{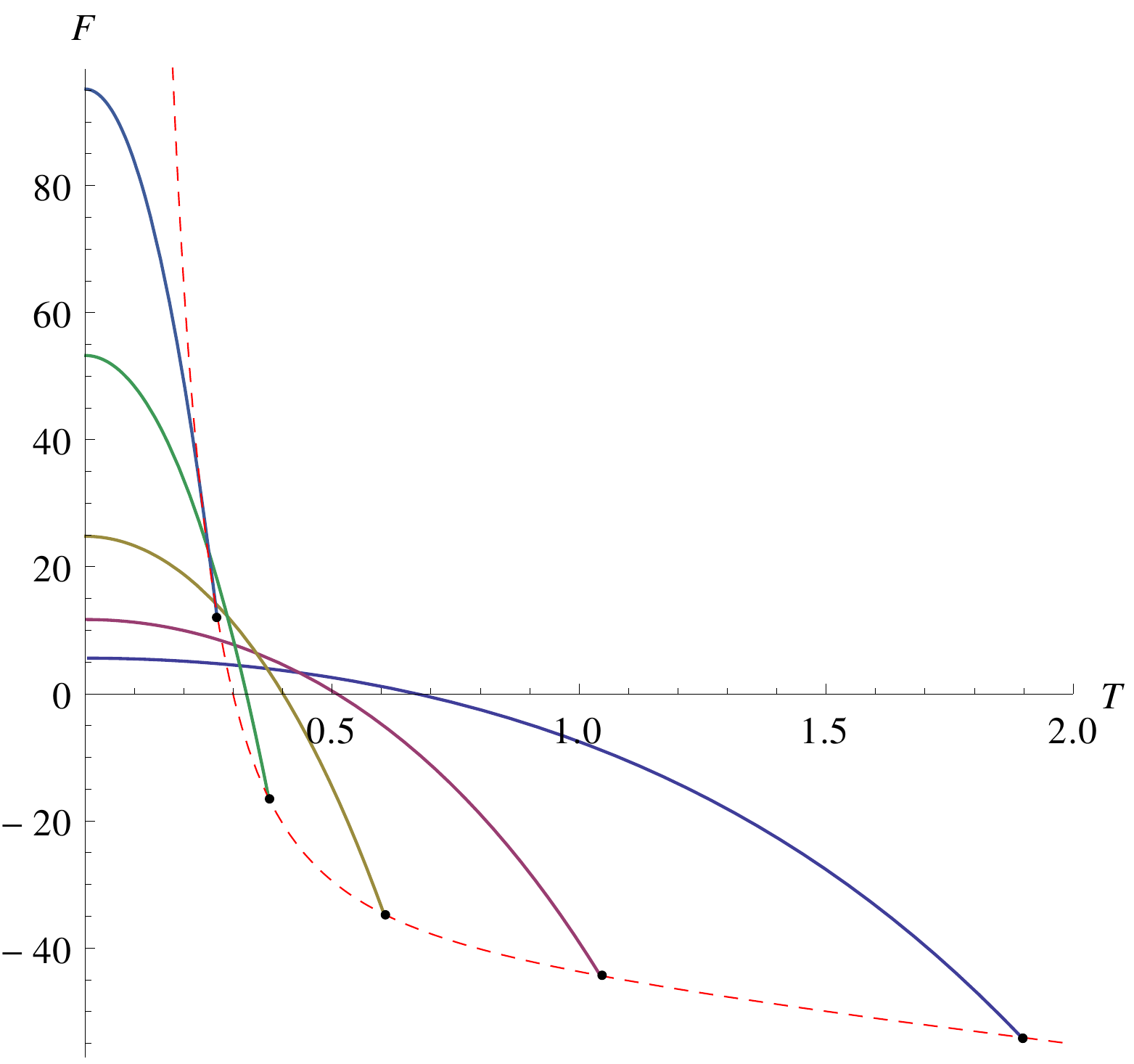}
\end{center}
\caption{Free energy of the bubble configuration, as compared to the thermal vacuum $\Lambda_+$ ($F=0$), as a function of the temperature $T = \beta_+^{-1}$ for several values of the LGB coupling, $\lambda = 0.1, 0.2, 0.4, 0.8, 1.35$ (from bottom to top in $y$-intercept), in five dimensions ($L=1$). The dashed line indicated the locus of the points where the different curves end (black dots), whose threshold temperature is called $T_\star$.} 
\label{Freefig}
\end{figure}

We can now plot the free energy as a function of the temperature and analyze the global and local stability of the different solutions (see Figure \ref{Freefig}).
For each value of the temperature, the saddle with the lowest value of the free energy will be the dominant one. The first thing we notice is that the thermalon configuration exists only for a limited range of temperatures for each $\lambda$. Above some threshold temperature, $T_\star$, which decreases with $\lambda$, the thermal AdS vacuum would be the only available static solution, thus being in this respect stable. A caution remark is necessary at this point since we shall remind the reader that in this toy model the AdS vacuum $\Lambda_+$ is BD unstable. This may not be the case in higher-curvature Lovelock gravities where analogous thermal phase transitions from AdS vacua free of BD instabilities to a dS black hole geometry should be possible. Besides, we would expect on general grounds weakly turbulent (nonlinear) instabilities of the form first discussed in \cite{turbulent}.

A comment is in order at this point. The phase diagram that corresponds to the free energy, $F$, seems to entail the occurrence of so-called {\it reentrant phase transitions},\footnote{We thank David Kubiznak and Robert Mann for pointing out this interesting possibility to us.} a familiar phenomenon in chemical physics that was earlier observed in the context of black hole thermodynamics \cite{RPT1,RPT2}. There is, however, an important subtlety hidden in the fact that Figure \ref{Freefig} is not comparing two different stable thermodynamical configurations. Instead, one of them --the {\it thermalon}-- is a finite temperature instanton describing an intermediate state, a bubble of true-vacuum that grows after popping-up, filling space with the new phase that changes the asymptotics to dS. Once this happens, we cannot refer any longer to the same diagram. On the other hand, as discussed shortly after Figure \ref{thermalonfig}, there are no regular thermalons with dS asymptotics. Therefore, we cannot reverse the process by changing the temperature, as required by reentrant phase transitions.

\begin{figure}[ht]
\begin{center}
\includegraphics[width=0.58\textwidth]{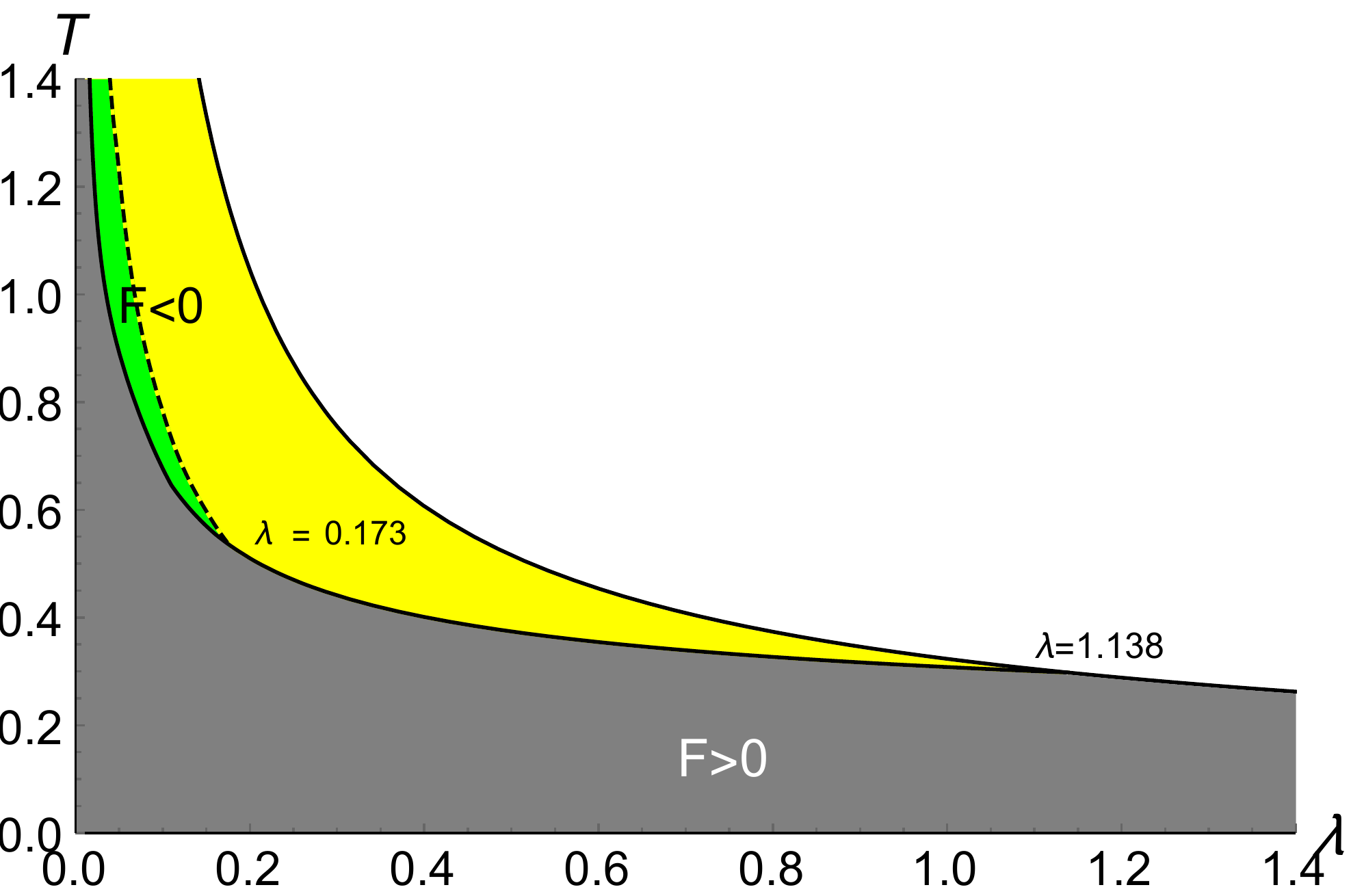}
\end{center}
\caption{Phase diagram, where T stands for $\beta_+^{-1}$ and $\lambda$ for the LGB coupling. The upper (respectively lower) solid curve corresponds to $T_\star(\lambda)$ (respectively, $T_c(\lambda)$). Both merge at $\lambda_\star = 1.13821$ (in five dimensions, $L=1$). Beyond $\lambda_\star$ there is no thermalon configuration. Between both lines there is the region where the thermalon free energy is negative and the transition is favored. In grey at the bottom we have the region of positive free energy of the thermalon. The dashed line separates the (upper--yellow/lower--green) regions with negative/positive black hole's specific heat. In white at the top the region where no thermalon is dynamically possible.}
\label{phasediagram}
\end{figure}

Maximal temperature for the thermalon happens when the mass of the interior spherically symmetric black hole with dS asymptotics reaches its upper bound given by the Nariai threshold \cite{bestiary}. The black hole with dS asymptotics inside the bubble may have negative specific heat when measured by the inner parameters $M_-$ and $\beta_-$ (see Figure \ref{phasediagram}; dashed line).
However, this is not problematic because it is not the relevant quantity in the thermodynamical description of the system. In fact, when analyzing the behavior of $M_+$ and $\beta_+$, as observed in the asymptotic region, the specific heat of the whole system --in the temperature range for which the thermalon exists-- turns out to be always positive. This is easily seen from the negative convexity of its free energy, $F(T)$. Therefore, we can say that the presence of the bubble {\it stabilizes} the black hole.

Let us focus in the five dimensional case. Regarding the sign of the free energy, we can distinguish two different regimes depending on the value of the LGB coupling (see Figure \ref{phasediagram}). For small enough values, $\lambda < \lambda_\star = 1.13821$, the free energy of the thermalon is positive at low temperatures, while it changes sign as we go to higher temperatures. For higher values of $\lambda$, on the other hand, the free energy remains positive for the whole range of temperatures of the thermalon. In the latter case, then, the dominant saddle is always the $\Lambda_+$ vacuum, whereas in the former it is so except for a limited range of intermediate temperatures, $T \in (T_c, T_\star)$. In this range the thermalon is the dominant saddle and the nucleation process is favored. When reaching the critical temperature the thermalon will form, but it will not remain in equilibrium for long. Even though thermodynamically stable, this configuration is dynamically unstable. Our bubble sits at a maximum of the potential and it eventually expands reaching the asymptotic region in finite proper time, thus changing the effective cosmological constant of the whole spacetime.

We effectively jump from one branch of solutions to another. This is important because the final (dS) branch does not suffer from the pathologies characterizing the unstable LGB AdS branch that was our initial asymptotics. Notice that ({\it quantum}) time scale for evaporation or thermal instability of the dS black hole is much larger than the {\it classical} expansion time, the former is order $\hbar^{-1}$ as compared to the latter. The ranges of temperatures for which each of the saddles is dominant can be represented in a diagram in terms of $\lambda$ as shown in Figure \ref{phasediagram}.

We said that for low enough temperatures the dominant saddle is always the $\Lambda_+$ vacuum. This does not mean that the transition does not take place. The probability of bubble formation in this situation is not zero but roughly $e^{-\hat{\mathcal{I}}}$ \cite{thermal1,thermal2,thermal3,thermal4,thermal5}. Waiting long enough the bubble will form and mediate a transition to the healthy branch of solutions. In this respect, we may say that the vacuum is metastable.

The only range of temperatures for which there is no transition is that of high temperatures for which the thermalon configuration does not even exist. The threshold temperature for this regime diverges as we approach small values of $\lambda$. At the same time, the free energy becomes lower and lower in the range in which it is positive. This means that the thermalon is more easily excited (the nucleation probability grows) as we go to lower values of $\lambda$, and that this configuration is available for all temperatures in the limit of very small $\lambda$. This is a strong indication that the AdS vacuum is more and more unstable as $\lambda \to 0$.

\section{Discussion and outlook}

We have described a novel scenario for transitions between AdS and dS asymptotics in higher-curvature gravity, without involving any additional matter fields. The phenomenon discussed in the present article is expected to take place in any higher-curvature theory as long as AdS and dS branches are allowed at the same time. The framework described here is general enough as to accommodate any series of higher-curvature terms, we would only need to adequately generalize the junction conditions.

For illustrative purposes we have chosen LGB gravity as the minimal model where this kind of transition takes place. For this theory we have all the necessary ingredients without the usual complications of higher-derivative terms. In particular, we have explicit junction conditions (given by {\it momenta} conservation in the Hamiltonian formalism) that enable us to construct the thermalon configuration. Nonetheless, this example has to be taken with a grain of salt, as a simple toy model, given that the theory has well known issues; namely, it generically violates causality\footnote{This signals the existence of an infinite tower of higher-spin particles whose presence does not necessarily affect the gravitational phase transitions discussed in this letter.} \cite{CEMZ}.

We are also truncating the effective action and analyzing the regime where the higher-curvature corrections become of the same order as the Einstein-Hilbert term, precisely where the rest of the higher-curvature series becomes relevant as well. In particular, the unstable vacuum curvature diverges as we take the LGB coupling to zero. All these concerns would be dealt with in a brane setup in string theory, where a consistent field theory limit is taken. In that respect, the LGB action is a promising candidate given that it arises both in heterotic string theory and in type II superstrings in the presence of wrapped probe D-branes.

This model might have as well interesting applications in the context of the gauge/gravity duality. In addition to the well known AdS/CFT correspondence there is a proposal for a dS/CFT duality \cite{StrdSCFT}. The transition mechanism presented in this letter suggests a possible avenue for a deeper understanding of both formulations by means of a framework where both types of asymptotics appear on equal footing in a given gravitational theory. Notice, nonetheless, that in all cases studied so far one of the vacua is unstable for some reason and we can always argue that we have to stick to the other asymptotics. There may be more general cases of transitions involving two perfectly healthy vacua in Lovelock theory.

The approach considered here is very different to other AdS/CFT descriptions of similar transitions (see, for instance, \cite{FHMMRS}) where the boundary in which the CFT lives is always unchanged, the transition taking place in the other asymptotic region of an eternal black hole. Our description is completely insensitive to this {\it other side} and even if we try to describe it, the same transition would happen there as well.

In asymptotically AdS spacetimes, boundary conditions are of paramount importance. In the case of transitions between two AdS vacua of different radii, one may argue that fixing the asymptotics (by means of reflecting boundary conditions) would make the bubble bounce back and collapse. In the case of AdS to dS transitions, though, this is actually impossible since the formation of a dS horizon makes the expansion of the bubble irreversible from that moment onwards \cite{CEGG2}.

This paper being a toy model, it seems clear that there are many avenues to improve our results. Addition of matter seems a necessary step towards embedding it in string theory.

\section*{Acknowledgements}

%
We are pleased to thank Gast\'on Giribet for early discussions on this project.
The work of JDE and JASG is supported in part by MINECO (FPA2011-22594), Xunta de Galicia (GRC2013-024), the Spanish Consolider-Ingenio 2010 Programme CPAN (CSD2007-00042) and FEDER.
AG and JDE gratefully acknowledges support from FONDECYT (Chile) grant 1141309.
JDE also wishes to acknowledge support from Marie Curie-IRSES/EPLANET, European Particle Physics Latin American Network, European Union 7th Framework Program, PIRSES-2009-GA-246806, and is thankful to the University of Buenos Aires, PUC Chile and UNAM Mexico for hospitality.
AG wishes to thank the Theory Group of Santiago de Compostela for their hospitality.
JASG acknowledges support from the Spanish FPI fellowship from FEDER grant FPA2011-22594. 



\begin{thebibliography}{99}

\bibitem{scalar1}
A.~D.~Linde,
``Is the Lee constant a cosmological constant?,''
JETP Lett.\  {\bf 19}, 183 (1974) [Pisma Zh.\ Eksp.\ Teor.\ Fiz.\  {\bf 19}, 320 (1974)].

\bibitem{scalar2}
M.~J.~G.~Veltman,
``Cosmology and the Higgs mechanism,''
Phys.\ Rev.\ Lett.\  {\bf 34}, 777 (1975).
  
\bibitem{3form1}
A.~Aurilia, H.~Nicolai and P.~K.~Townsend,
``Hidden constants: the theta parameter of QCD and the cosmological constant of N=8 supergravity,''
Nucl.\ Phys.\ B {\bf 176}, 509 (1980).

\bibitem{3form2}
M.~J.~Duff and P.~van Nieuwenhuizen,
``Quantum inequivalence of different field representations,''
Phys.\ Lett.\ B {\bf 94}, 179 (1980).

\bibitem{3form3}
P.~G.~Freund and M.~A.~Rubin,
``Dynamics of dimensional reduction,''
Phys.\ Lett.\ B {\bf 97}, 233 (1980).

\bibitem{3form4}
S.~W.~Hawking,
``The cosmological constant is probably zero,''
Phys.\ Lett.\ B {\bf 134}, 403 (1984).

\bibitem{3form5}
M.~Henneaux and C.~Teitelboim,
``The cosmological constant as a canonical variable,''
Phys.\ Lett.\ B {\bf 143}, 415 (1984).

\bibitem{instanton1}
S.~R.~Coleman and F.~De~Luccia.
``Gravitational effects on and of vacuum decay,''
Phys.\ Rev.\ D {\bf 21}, 3305 (1980).

\bibitem{instanton2}
J.~D.~Brown and C.~Teitelboim,
``Dynamical neutralization of the cosmological constant,''
Phys.\ Lett.\ B {\bf 195}, 177 (1987).

\bibitem{instanton3}
J.~D.~Brown and C.~Teitelboim,
``Neutralization of the cosmological constant by membrane creation,''
Nucl.\ Phys.\ B {\bf 297}, 787 (1988).

\bibitem{thermal1}
J.~S.~Langer,
``Statistical theory of the decay of metastable states,''
Ann. Phys.\ (N.Y.) {\bf 54}, 258 (1969).

\bibitem{thermal2}
A.~D.~Linde,
``On the vacuum instability and the Higgs meson mass,''
Phys.\ Lett.\ B {\bf 70}, 306 (1977).

\bibitem{thermal3}
A.~D.~Linde,
``Fate of the false vacuum at finite temperature: theory and applications,''
Phys.\ Lett.\ B {\bf 100}, 37 (1981).

\bibitem{thermal4}
A.~D.~Linde,
``Decay of the false vacuum at finite temperature,''
Nucl.\ Phys.\ B {\bf 216}, 421 (1983) [Erratum-ibid.\ B {\bf 223}, 544 (1983)].

\bibitem{thermal5}
I.~Affleck,
``Quantum statistical metastability,''
Phys.\ Rev.\ Lett.\  {\bf 46}, 388 (1981).

\bibitem{thermalon} 
A.~Gomberoff, M.~Henneaux, C.~Teitelboim and F.~Wilczek,
``Thermal decay of the cosmological constant into black holes,''
Phys.\ Rev.\ D {\bf 69}, 083520 (2004) [hep-th/0311011].

\bibitem{adsds} 
W.~Kim and M.~Yoon,
``Transition from AdS universe to dS universe in the BPP model,''
JHEP {\bf 0704}, 098 (2007) [gr-qc/0703019].

\bibitem{GuptSingh} 
B.~Gupt and P.~Singh,
``Non-singular AdS-dS transitions in a landscape scenario,''
Phys.\ Rev.\ D {\bf 89}, 063520 (2014) [arXiv:1309.2732 [hep-th]].

\bibitem{Odintsov1} 
M.~Cvetic, S.~Nojiri and S.~D.~Odintsov,
``Black hole thermodynamics and negative entropy in de Sitter and anti-de Sitter Einstein-Gauss-Bonnet gravity,''
Nucl.\ Phys.\ B {\bf 628}, 295 (2002) [hep-th/0112045].

\bibitem{Odintsov2} 
S.~Nojiri and S.~D.~Odintsov,
``The de Sitter/anti-de Sitter black holes phase transition?,''
gr-qc/0112066.

\bibitem{BD}
D.~G.~Boulware and S.~Deser,
``String generated gravity models,''
Phys.\ Rev.\ Lett.\  {\bf 55}, 2656 (1985).

\bibitem{Wheeler1}
J.~T.~Wheeler,
``Symmetric solutions to the Gauss-Bonnet extended Einstein equations,''
Nucl.\ Phys.\ B {\bf 268}, 737 (1986).

\bibitem{Wheeler2}
J.~T.~Wheeler,
``Symmetric solutions to the maximally {Gauss-Bonnet} extended Einstein equations,''
Nucl.\ Phys.\ B {\bf 273}, 732 (1986).

\bibitem{bestiary}
X.~O.~Camanho and J.~D.~Edelstein,
``A Lovelock black hole bestiary,''
Class.\ Quant.\ Grav.\  {\bf 30}, 035009 (2013) [arXiv:1103.3669 [hep-th]].

\bibitem{CESdS} 
X.~O.~Camanho, J.~D.~Edelstein and J.~M.~S‡nchez De Santos,
``Lovelock theory and the AdS/CFT correspondence,''
Gen.\ Rel.\ Grav.\  {\bf 46}, 1637 (2014) [arXiv:1309.6483 [hep-th]].

\bibitem{CEGG1}
X.~O.~Camanho, J.~D.~Edelstein, G.~Giribet and A.~Gomberoff,
``New type of phase transition in gravitational theories,''
Phys.\ Rev.\ D {\bf 86}, 124048 (2012) [arXiv:1204.6737 [hep-th]].

\bibitem{CEGG2}
 X.~O.~Camanho, J.~D.~Edelstein, G.~Giribet and A.~Gomberoff,
``Generalized phase transitions in Lovelock gravity,''
Phys.\ Rev.\ D {\bf 90} (2014) 6,  064028 [arXiv:1311.6768 [hep-th]].

\bibitem{quenches}
A.~Buchel, R.~C.~Myers and A.~van Niekerk,
``Nonlocal probes of thermalization in holographic quenches with spectral methods,''
JHEP {\bf 1502}, 017 (2015) [arXiv:1410.6201 [hep-th]].

\bibitem{CaiGuo}
R.~-G.~Cai and Q.~Guo,
``Gauss-Bonnet black holes in dS spaces,''
Phys.\ Rev.\ D {\bf 69}, 104025 (2004) [hep-th/0311020].

\bibitem{Davis2003}
S.~Davis,
``Generalized Israel junction conditions for a Gauss-Bonnet brane world,''
Phys.\ Rev.\ D {\bf 67}, 024030 (2003) [arXiv:0208205 [hep-th].

\bibitem{wormholes}
E.~Gravanis and S.~Willison,
``Mass without mass: from thin shells in Gauss-Bonnet gravity,''
Phys.\ Rev.\ D {\bf 75}, 084025 (2007) [arXiv:0701152 [gr-qc].

\bibitem{turbulent}
P.~Bizon and A.~Rostworowski,
``On weakly turbulent instability of anti-de Sitter space,''
Phys.\ Rev.\ Lett.\  {\bf 107}, 031102 (2011) [arXiv:1104.3702 [gr-qc]].

\bibitem{RPT1}
N.~Altamirano, D.~Kubiznak and R.~B.~Mann,
``Reentrant phase transitions in rotating anti--de Sitter black holes,''
Phys.\ Rev.\ D {\bf 88}, 101502 (2013) [arXiv:1306.5756 [hep-th]].
  
\bibitem{RPT2}
A.~M.~Frassino, D.~Kubiznak, R.~B.~Mann and F.~Simovic,
``Multiple reentrant phase transitions and triple points in Lovelock thermodynamics,''
JHEP {\bf 1409}, 080 (2014) [arXiv:1406.7015 [hep-th]].

\bibitem{CEMZ}
X.~O.~Camanho, J.~D.~Edelstein, J.~Maldacena and A.~Zhiboedov,
``Causality constraints on corrections to the graviton three-point coupling,''
arXiv:1407.5597 [hep-th].

\bibitem{StrdSCFT}
A.~Strominger,
``The dS/CFT correspondence,''
JHEP {\bf 0110}, 034 (2001) [hep-th/0106113].

\bibitem{FHMMRS}
B.~Freivogel, V.~E.~Hubeny, A.~Maloney, R.~C.~Myers, M.~Rangamani and S.~Shenker,
``Inflation in AdS/CFT,''
JHEP {\bf 0603}, 007 (2006) [hep-th/0510046].

\end{thebibliography}
\end{document}